# HOLISTIC EVALUATION OF XML QUERIES WITH STRUCTURAL PREFERENCES ON AN ANNOTATED STRONG DATAGUIDE

Maurice TCHOUPÉ TCHENDJI, Adolphe Gaius NKUEFONE, and Thomas TÉBOUGANG TCHENDJI

Department of Mathematics and Computer Science, Faculty of Sciences, University of Dschang, PO Box 67, Dschang Cameroon

*ABSTRACT*

*With the emergence of XML as de facto format for storing and exchanging information over the Internet, the search for ever more innovative and effective techniques for their querying is a major and current concern of the XML database community. Several studies carried out to help solve this problem are mostly oriented towards the evaluation of so-called exact queries which, unfortunately, are likely (especially in the case of semi-structured documents) to yield abundant results (in the case of vague queries) or empty results (in the case of very precise queries). From the observation that users who make requests are not necessarily interested in all possible solutions, but rather in those that are closest to their needs, an important field of research has been opened on the evaluation of preferences queries. In this paper, we propose an approach for the evaluation of such queries, in case the preferences concern the structure of the document. The solution investigated revolves around the proposal of an evaluation plan in three phases: rewriting-evaluation-merge. The rewriting phase makes it possible to obtain, from a partitioning-transformation operation of the initial query, a hierarchical set of preferences path queries which are holistically evaluated in the second phase by an instrumented version of the algorithm TwigStack. The merge phase is the synthesis of the best results.*

## 1. INTRODUCTION

A semi-structured XML document is an electronic document whose textual content has a certain regularity (*well formed document*), and has a structure that is not constrained by a model: its structure is flexible [1]. Textual content and structural flexibility make semi-structured documents excellent candidates for the exchange and storage of data: we talk of XML Databases (XML DBs).

The data stored in the DBs are generally exploited through a dedicated query language (SQL for RDB - Relationnal Database -, XPath, XQuery for XML DBs, etc) allowing the user to express queries to be efficiently evaluated by a query engine. The quest for more innovative and effective techniques for evaluating user queries is becoming a major and current concern of the XML database community. Several authors have looked into it and have proposed solutions that initially consisted in the rewriting of XML queries into SQL ones, and then, in a second step, the proposal of native evaluation algorithms [2, 3, 4] for the so-called *exact* or *strict queries*, which, unfortunately, are likely (especially for the case of semi-structured documents) to return abundant results (case of vague queries) or empty results (case of very specific queries).

From the observation that users making queries are not necessarily interested in all possible responses to their queries, but rather those that come closest to their needs, an important field of research have been opened on the evaluation of *queries with preferences*: these queries consist of two parts, specifying mandatory requirements called constraints and optional requirements called *wishes* or *preferences* [5, 6]. It is understood that a response to a query with preference must imperatively satisfy the first part and possibly the second; however, if there is even a satisfactory response to both the first and the second part, only the responses that satisfy both must be returned as a result.

The problem addressed in this paper concerns the evaluation of XML queries with structural preferences (preferences relating to the structure of the document). Several approaches to evaluate such queries have already been proposed in the literature (see section 2.2). They usually differ by the document indexing technique used (like DataGuide [7] or region encoding [8]) or by their evaluation strategy (like rewriting queries before their evaluation, [9, 7, 8] or direct evaluation [10]).

We propose in this paper an approach to evaluate queries with *structural preferences*. The investigated solution concerns the case of documents indexed by a DataGuide [11] annotated by the region encoding [12, 2, 4]; it revolves around the proposal of the preferences queries evaluation strategy in three phases: rewriting-evaluation-merge. The rewriting phase makes it possible to obtain (this is the first main contribution of the paper), from a partitioning-transformation operation of the initial query, a hierarchical set of preference path queries which is (this is the second main contribution of the paper) holistically evaluated in the evaluation phase by an instrumented version of the *TwigStack* algorithm [2]. The merging phase consists in synthesizing the best results by making use of a so-called *preferenceTable* in order to store the candidate results, from which the best ones (i.e. the un-dominated ones) will be selected by means of the Skyline operator [13]. A synoptic view of the proposed evaluation approach is schematized in the figure 1.

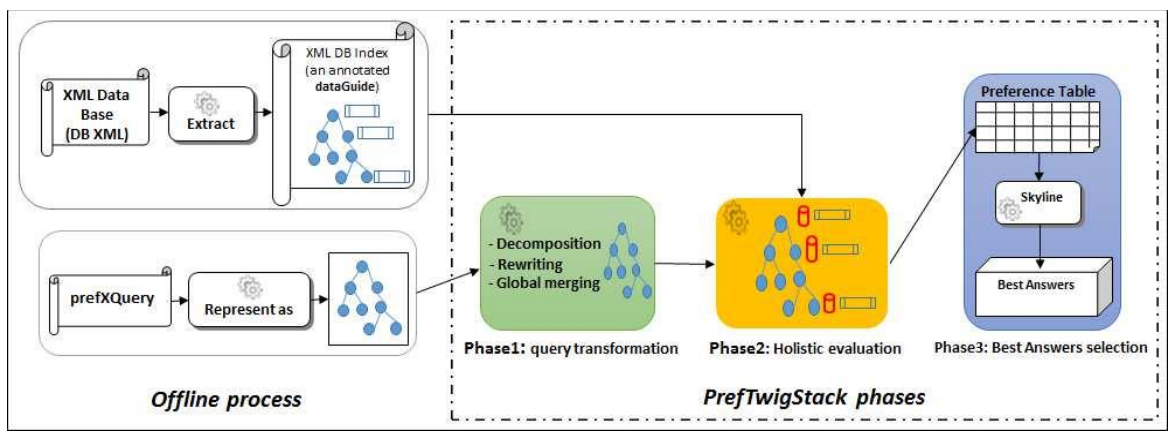

Figure 1. A synoptic view of our holistic evaluation approach of preference queries

**Organization of the manuscript:** section 2 introduces the concepts of XML documents (representation, indexing) and their queries, while section 3 is dedicated to the presentation and illustration of our holistic evaluation approach of preference queries. In section 4, we fully unfold a running example of *prefTwigStack* highlighting the majors concepts outlined in this paper. Finally, we conclude the paper in the section 5.

## 2. XML DOCUMENTS AND QUERIES

### 2.1. Representations And Indexing Of XML Documents

An *abstract XML document*[1] is represented by a labeled tree $D =(Nd, Ed)$ where $Nd$ is a set of labeled nodes, $Ed$ a set of arcs each connecting two nodes of $Nd$. For any node $x \in Nd$ the function $label_d(x)$ returns its label.

XML documents are generally exploited through indexes that can be grouped into three categories [4]: *path indexes*, which is a summary of all paths from the root to any node of the considered XML document, *node indexes*, and the category of *sequence based indexes*. A detailed presentation of those commonly used is given in [14]. In this paper, the index used is the one that is derived from the hybridization of a path index: the (strong) DataGuide [11] and a index based on the region encoding [12]; it is this index which is used for the implementation of TwigX-Guide [4].

Let's recall that, in its original form, a DataGuide is a summary of all the paths contained in a document in such a way that, every path of the source document appears exactly once in the DataGuide. The DataGuide is usually smaller than the document whose it is the index and can be used to answer very effectively path queries containing only *ParentChild* (P-C) relationships. However, it is unusable for the evaluation of queries containing the *Ancestor-Descendant* (A-D) relations because, it does not retain in its representation, the hierarchical structural relations that exist between the nodes of the document which it represents [11, 15]. Nevertheless, if we annotate a DataGuide for example with the nodes region encoding as in TwigX-Guide [4], we can couple it to a query rewriting strategy to effectively evaluate twig queries. Let's recall also that in the indexing based on nodes region encoding [12], each document node is represented by a triplet *(Start, End, Level)*: *Start* and *End* represent respectively the start and end positions of the element in the document; *Level* is the depth of the element in the tree representation of the document. With this convention, as in [3], the index of a document consists of a set of sorted linked lists $T_a$ of occurrences of nodes of type $a$. These lists are sorted according to the component *Start* of the triplet.

Note that, for any couple of nodes a and b respectively represented by the triplets *(start_a, end_a, level_a)* and *(start_b, end_b, level_b)*, one of the advantages of this representation is that it allows to determine the relations (P-C) or (A-D) in constant time. Indeed, $a$ is an ancestor of $b$ if and only if $start_a < start_b < end_a$. If in addition $level_a + 1 = level_b$, then $a$ is the *parent* of $b$. Subsequently, we will say of a node $a$ that it *covers* a node $b$ if $b$ is a descendant of $a$.

### 2.2. XML Queries: Representation And Evaluations

Like a classic DB, an XML document contains information (data). It also encapsulates a structure that must be taken into account when querying. Thus, an XML query concerns not only the content (the data) but also the structure (the structural relations that the user wishes to have between the different occurrences of the elements).

As for the documents representation, an XML query $Q$ can be represented by a tree $Q = (Nq, Eq)$ in which $Nq$ is a set of labeled nodes, $Eq$ a set of arcs each linking two nodes of $Nq$. In $Eq$ there are two types of arcs: those linking a *parent node* to a *child node* denoted $x/y$ and those linking a node to one of his descendants noted $x//y$. For any node $x \in Nq$ the function $label_q(x)$ returns its label, and we write $\Sigma = \{labelq(x), x \in Nq\}$ the set of all labels of the query $Q$.

---

[1] In an abstract XML document, text and attribute nodes are ignored: these are not of any interest for the purely structural treatments that interest us in this paper.

Let, $Q = (Nq, Eq)$ be a query, $D = (Nd, Ed)$ an XML document, two nodes $nd \in Nd$ and $nq \in Nq$: we'll say that $nd$ is an *instance* or an *occurrence* of $nq$ in $D$ if $labelq(nq) = labeld(nd)$. Matching the query $Q$ on the document $D$ consists in finding for every node of $Q$, their occurrences in $D$ and such that, these occurrences satisfy the same structural constraints than those defined between the nodes $Nq$. If it is the case, we say we found a match of $Q$ in $D$. Evaluating a query $Q$ on a document $D$ is equivalent to find all its matches in $D$.

Several techniques for evaluating XML queries exist in the literature. Some (older) of them rewrite XML queries to another format, usually the SQL format; others, on the other hand, operate natively. In the second category, there are two main approaches of evaluation [4]: one carrying the depth first search (DFS) tree traversal of the tree representation of the document to match the query tree [3, 9, 8], and the one proceeding by *decomposition-matching-merge* [2, 4]. The first algorithms proposed in the second approach proceeded by binary decomposition of the query nodes, with the main disadvantage that, they construct many intermediate results that are not necessarily used for the production of the final solution. Moreover, these algorithms perform many mergers that explode their temporal complexities. New techniques have been developed in the wake of the second approach to address this limitation. The most famous use a holistic evaluation approach originally proposed by Bruno et al. [2], and developed in many works [16, 4].

### 2.3. Preference queries

When the query matching is made in such a way that each query node is absolutely associated with at least one document node [2,3,4,9], the corresponding query is said to be an *exact query*. There are also queries with so-called *preferences* nodes also called *preference queries*. A matching of such a query may not contain occurrences of the preference nodes but, necessarily, those of the so-called *strict* or *exact nodes* [6,7,8,10]. Evaluating a preference query on a document amounts to returning the best matches of it. In order to do this, we first return exclusively only the matches containing all the occurrences of the query nodes and, if there is no such matching, we return those containing the "maximun" matching of preference nodes.

In the current study, the language used to express preference queries is the one proposed by Sara Cohen et al. [7]. A preference node is followed by the "!" symbol. For example, in the preference query $Q1 = A[B!\ [C/D]/E]/F[G[//H[I!/J]/K]/L]/M$ the nodes "$B$" and "$I$" are preference nodes. A query path in which a preference node can only appear as the last node is called a *prefPathQuery*. They are on the form $l1/.../lm/l(m+1)!$; $Q2 = A/B!$ is an example.

## 3. HOLISTIC EVALUATION OF PREFERENCES QUERIES

We present in this section *PrefTwigStack*, a holistic evaluation algorithm of XML preference queries. It is decomposed into three routines corresponding to as many processing phases: rewriting-evaluation-merge (see figure 1).

### 3.1. Preftwigstack: The Rewriting Phase Of The Initial Query

The purpose of the rewriting phase is to transform a preference query into an equivalent one in which nodes are *prefPathQuery*: the obtained query is then easily evaluable on a DataGuide. This phase consists of three sub-phases: local decomposition - rewriting - global merging.

### 3.1.1. The Local Decomposition Sub-Phase

The initial query $t$ is decomposed into a set of subqueries $t1,..., tn$ forming a (tree) partition $\Pi t$ [2] of $t$ and such that, for any subquery $ti$ (it is a subtree), it can not exist a (A-D) relation between its nodes, but also, it can only have preference nodes at its leaf. $\Pi t$ is constructed as follows:

   a) find the largest prefix $t1$ of $t$ (it is a subtree) such that its leaf nodes which are either leaves in $t$, or preference nodes or else, are nodes linked to another node of $t$ by an (AD) relation. In order to do this, during the $t$'s traversal, we cut any arc starting from a preference node or any arc starting from a node having an arc representing an (A-D) relation.
   b) prune $t1$ from $t$: it is the first element of the $\Pi t$ partition; in fact, it is the root element of (the tree) $\Pi t$.
   c) if following the pruning of $t1$ from $t$, the residual forest is not empty, then, each subtree of this one is pruned equally (as in a) and in b)); each time, the obtained prefix is added as the son of a $\Pi t$ node, at the same relative position like the one occupied by the $t$'s currently pruned subtree in $t$[3]. This process is repeated until an empty forest is obtained (fig. 2.a).

### 3.1.2. The Local Rewriting Sub-Phase

The goal here is to rewrite a twig subquery[4] (the tag of a $\Pi t$ node), in an equivalent one whose nodes are tagged with *prefPathQuery* (figure 2.b).

The rewriting technique is inspired by the one used by Su-Cheng et al. [4] for the evaluation of exact queries on a DataGuide. Note that, the rewriting approach that they have developed, significantly reduces the number of subqueries to be evaluated on DataGuide and consequently, the number of joins to perform in order to synthesize the final solution.

The rewrite approach described in [4] uses the *topBranchNode* notion which, for a given $t$ twig query, designates the first twig node that is encountered on the path stemming from the $t$'s root. Following this notion, we introduce that of *topBranchNodePath* which designates the path going from the $t$'s root to its *topBranchNode*.

Let $q$ be a twig (sub)query produced by the previous phase, and $ti$ be a subtree (ie, a subquery) of $q$, let's consider *absPathTo_ti* as the absolute path to the root node of $ti$ in $q$. The main idea of our rewriting technique is: a) to determine the *topBranchNodePath_ti* of $ti$; it will be used to build *absPathToTopBranch-NodePath_ti* which will be the root node label of the tree to be generated by the $ti$ rewriting, and the prefix for the tags of the other nodes of that tree: *absPathToTopBranchNodePath_ti = absPathTo_ti/topBranchNodePath_ti* b) to recursively transform any subtree $st$ of the $ti$ *topBranchNode* into a tree tagged by pathQuery. To do this, considering that *absPathToTopBranchNodePath_ti* is the absolute path to the *topBranchNode* of *ti*, if *st* is a path query, then, add it as the child node of the node labeled by *absPathToTopToBranchNodePath_ti/st* at the same position as *st* was in *ti*; this node will be tagged *absPathToTopBranchNodePath_ti/st*. Otherwise, recursively call the function *rewrite*[5] with *absPathToTopBranchNodePath_ti* and the subtree *st* as effective parameters. The recursive call result will be added as the child node of the node labeled by the *absPathToTopBranchNodePath_ti* at the same position as *st* was in *ti*.

---

[2] The $\Pi t$ partition is represented as a tree in which each node contains a subquery $ti$ of $t$. This tree representation of the partitioning of $t$, makes it possible to preserve in the subqueries $ti$ all the hierarchical relations (P-C, A-D, etc ) that existed between $t$'s nodes.

[3] In order to make the processing of the *global merge* sub-phase possible, each node of the tree encoding the partitioning of $t$ is decorated by a triplet (*leafNum*, *relPos*, *relType*) such that, if the subquery $tic$ of $\Pi t$ has as parent the subquery $tip$ in $\Pi t$, *relPos* is the relative position of the root node of $tic$ in the list of child nodes of *nelag* (*nelag* is the $tip$'s leaf node, whose $tic$'s root node is one of his children: *nelag* is either a preference node, or a node having an (A-D) relation whose pruning led to the creation of $tic$), and *typeRel* is the relationship type ((A-D) or (P-C)) that links *nelag* to the root node of $tic$ in $t$. *leafNum* is the *nelag*'s relative position among the leaves of $tip$.

[4] Recall that, it must at the best of times have preference nodes exclusively at his leaves.

### 3.1.3. The global merging sub-phase

The purpose of this subphase is to produce a query whose nodes are tagged by preference path queries by merging the annotated rewriting of subqueries from the previous subphase. Any *tic* rewriting of a subquery *t* annotated with (*leafNum*, *relPos*, *typeRel*), which parent is the rewritten subquery *tip*, is inserted as *relPos*$^{th}$ subtree of the leaf number *leafNum* of *tip*; the relation linking it to this leaf being *typeRel*.

Figure 2 gives an overview of the results of the processing carried out in each of these subphases on the *Q1* query presented in section 2.3.

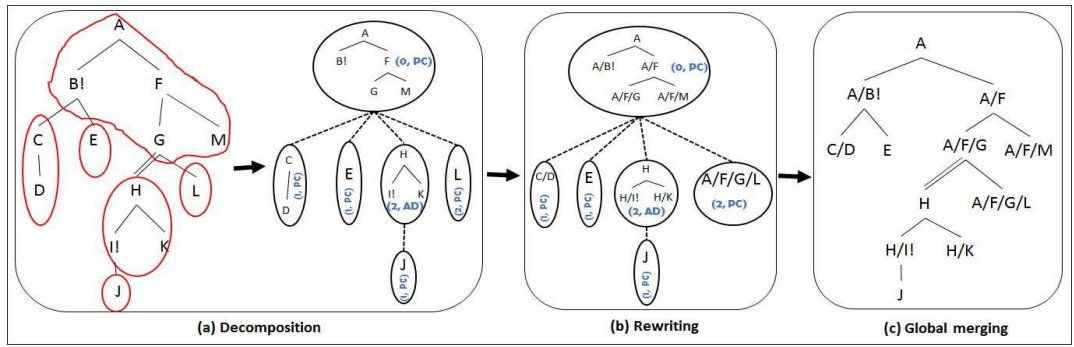

Figure 2. Overview of prefTwigStack rewriting phase on the initial query Q1

## 3.2. Preftwigstack: Holistic Evaluation Phase Of The Rewritten Query

This phase takes as input a query in which each node is decorated with a *prefPath-Query*. It aims to return all the solutions resulting from the holistic evaluation of the latter with production for any solution, information indicating the occurrences of the preference nodes which it eventually integrates; those informations are directly inserted in a so-called *prefTable* [8] a soon as they are produced.

The algorithm proposed for this phase is an instrumentalized version of the *TwigStack* algorithm [2] for the evaluation of a tree whose nodes are labeled with *prefPathQuery*[6].

Like in *TwigStack*, a stack and a list are associated with each node of the query tree. However, unlike *TwigStack* which for a query node already has the list of occurrences of his label via the document index, here, this list is obtained from the annotated DataGuide after the evaluation of the *prefPathQuery* who label the query node. For the evaluation itself, *TwigStack* [2] is used to holistically merge the lists associated with the different nodes. But, how are these lists obtained?

---

[5] The function *rewrite* allows to rewrite the tree taken as a second parameter into an equivalent tree whose nodes are labeled by path queries, all prefixed by the path that the function takes as its first parameter.
[6] Note that in *TwigStack*, nodes are labeled by element names and not by paths as they are here.

### 3.2.1. Evaluating A Prefpathquery On An Annotated Dataguide

A *prefPathQuery Chnq* tagging a given node of a query tree is either a strict path of the form $Chnqs = l1/.../lm$ or, a preference path of the form $Chnqp = l1/.../lm/l(m+1)!$ having $l(m+1)!$ as the single preference node. The evaluation of such a query on annotated DataGuide is immediate when it is a *strict path*. Whereas, for a preference path, it is necessary to take into account the cases where the occurrence of the preferred node may be absent.

In order to sharpen our intuition for a better understanding of the proposed evaluation technique, let's consider the evaluation of a simple example of a preference path query $Q3 = /A/B!/C$ and examine the different cases that we can face.

Consider that the evaluation of the strict query $q = /A$ (resp. $q' = /A/B$) on the DataGuide produced the list $TA =[a0, a0', a1, a2, a0'']$ (resp. $TB =[b1, b2]$), such that in the document, $a1$ (resp. $a2$) is the parent of $b1$ (resp. $b2$). Three possible cases represented in figure 3 are to be examined:

Case 1) processing of *C*'s occurrences that *precede*[7] the first occurrence $b1$ of $B$ and are part of the solution. Such occurrences can be either direct childs of an occurrence of $A$ preceding $a1$ in the list $TA$, or direct childs of $a1$ (figure 3.a). For the treatment of this case, by noting $nprec(a1)$ the number of $A$ occurrence that precede or are equal to $a1$ in $TA$, we must insert in the list $TB$, $nprec(a1)$ *pseudo-occurences* of $B$ noted $\varepsilon1 ... \varepsilon n_{prec(a1)}$ and considered as *pseudo-parents* of the $C$'s occurrences processed in the current case. For the case illustrated in the figure 3.a, we will stamp each $\varepsilon i$, $1 \leq i \leq 3$ respectively with the region encoding pairs *(nextL(a0), nextR(a0))*, *(nextL(a0'), nextL(a1))* and *(nextL(a1), nextL(b1))* where *nextL(ai)* is equal to the *start* component of $ai$ and *nextR(ai)* is equal to the *end* component of $ai$: the stamps calculation scheme is materialized by the green dashed lines in the figure 3.a.

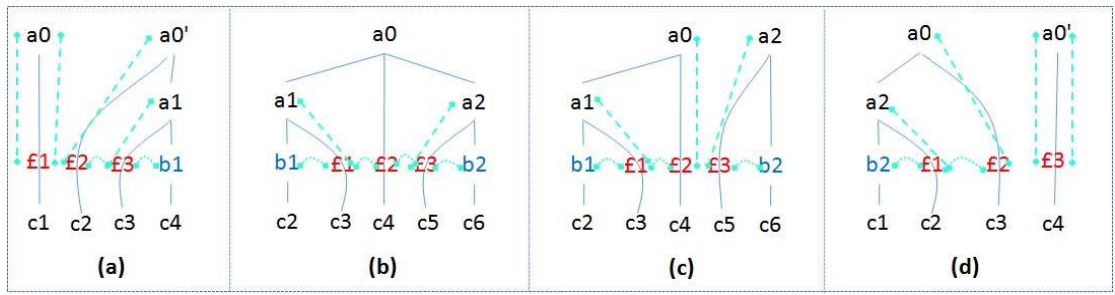

Figure 3. *Various cases according to the position of B's occurrences: (a) at the head of the TB list, (b) and (c) between two consecutives B's occurrences in TB list, (d) at the end of the TB list.*

Case 2) processing of *C*'s occurrences that *follows*[8] the last occurrence $b2$ of $B$ and are part of the solution. Such occurrences can be either direct childs of an occurrence of $A$ following $a2$ in $TA$, or direct childs of either $a2$ or an occurrence of $A$ covering $a2$(figure 3.d). Noting $n_{follow(a2)}$ the number of $A$'s occurrences following $a2$ or covering $a2$ or equal to $a2$ in $TA$, for the treatment of this case, we must insert in list $TB$, $n_{follow(a2)}$ *pseudo-occurences* of $B$ noted $\varepsilon1 ... \varepsilon n_{follow(a2)}$ and interpreted as *pseudo-parents* of any occurrences of $C$ processed in the current case. For the case illustrated in the figure 3.d, we will stamp each $\varepsilon i$, $1 \leq i \leq 3$ respectively with the region encoding pairs *(nextR(b2), nextR(a2))*, *(nextR(a2), nextR(a0))* and *(nextL(a0'), nextR(a0'))*.

Case 3) processing of *C*'s occurrences appearing between two successive occurrences $b1$ and $b2$

---

[7] It will be said that a node *a(starta, enda, levela)* precedes a node *b(startb, endb, levelb)* if ( $starta < startb$).

[8] It will be said that a node *a(starta, enda)* follows a node *b(startb, endb)* if ($endb < starta$).

of *TB*, and are part of the solution. We can notice that, here, $a1$ and $a2$ can be covered by the same occurrence of *A* (figure 3.b) or not (figure 3.c). However, in all cases, the number of *pseudo-occurrences* $n[a1−a2]$ of *B* to be inserted in *TB* is equal to the number of occurrence of *A* between $a1$ and $a2$ in *TA*: $n[a1−a2]= |ak(start_{ak}, end_{ak}), end_{a1} < end_{ak} < start_{a2}|$. For the case illustrated in the figure 3.b, we will stamp each $\varepsilon_i$, $1 \leq i \leq 3$ respectively with the region encoding pairs *(nextR(b1), nextR(a1))*, *(nextR(a1), nextL(a2))* and *(nextL(a2), nextL(b2))*. We deduce from the study of these three cases that the list *Tnq* of the occurrences which will be associated with a node labeled by a preference path query $Chnq = l1/.../lm/l(m+1)!$ can be built as follows: 1) evaluate the two strict queries $Chnq1 = l1/.../lm$ and $Chnq2 = l1/.../lm/l(m+1)$ on the DataGuide. Let's note $\alpha =[a0,..., am]$ and $\beta =[b1,..., bn]$ the corresponding answers[9]. 2) initialize *Tnq* to β: $Tnq =[b1,..., bn]$. 3) create and insert appropriately in *Tnq* the *pseudo occurrences* $\varepsilon_0,..., \varepsilon_m$ as shown above.

### 3.2.2. The holistic evaluation

The holistic evaluation phase of *PrefTwigStack* uses a very slightly adapted version of the *TwigStack* algorithm and, like the latter, it operates in two steps: the intermediate solutions are produced for all the root to leaf paths queries of the query tree; these are then merged to obtain the final answer.

The major difference between the algorithm used in the holistic evaluation phase of *PrefTwigStack* and *TwigStack* lies on the type of the nodes tags, and on how to get the lists associated with the query nodes; those lists are provided in *TwigStack* while they are built in the running time in *PrefTwigStack*. A few instructions must be add in the TwigStack merge function so that, each solution resulting from the merge is inserted into the *preferenceTable* table. Recall that, according to [8], during an insertion, we must indicate in a boolean way and for each preference node, if the solution being inserted includes or not one of its occurrences (see table 1).

### 3.3. PrefTwigStack: the best answers merging phase

In this phase, the answers recorded in the *preferenceTable* during the previous phase are filtered using *the skyline operator* [13, 8] to retain only those that are not *dominated* in the sense of the preference relationship.

Let's consider two tuples $p =(p1,..., pk, pk+1,... pn)$ and $q =(q1,..., qk, qk+1,... qn)$ of a relational table *R* whose schema is $R(P1, ..., PK, PK+1,..., Pn)$[10]. For a query in which the preferences are related to the fields $Pk+1,..., Pn$, by using *the skyline operator*, we will say that *p dominates q* and we write $p > q$, if the following three conditions are satisfied: (1) $pi = qi$, for all $i = 1, 2,... k$. (2) $pi \geq qi$ for all $i =(k + 1),..., n$. (3) there is $i$, $(k + 1) \leq i \leq n$, and $pi > qi$. The best answers will then be the non-dominated answers of the table *preferenceTable*.

## 4. AN ILLUSTRATION OF PREFTWIGSTACK

In this section, we show a running example of *prefTwigStack*. The query $Q4 = /A[B!/C]/D/E$ schematized in the figure 4.b is evaluated on the XML document represented in the figure 4.c. An index of this document is schematized in the figure 5. By applying the rewrite procedure presented in the section 3.1, we have in the figure 4.b a representation of the rewritten query.

---

[9] Recall that the $bi$, $i = 1,... n$ as well as the $ai$, $i = 1,..., m$ are actually triplets *(starti, endi, leveli)* used in region encoding.

[10] According to this study, these two tuples are two answers to a preference query and *R* represents the table *preferenceTable*.

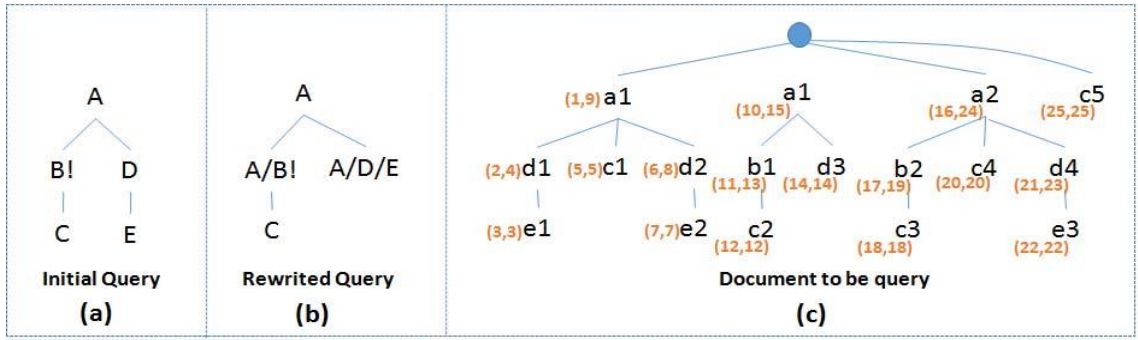

Figure 4. (a) An intial query, (b) A query after transformation, (c) An annotated (with (start, end)) document to be query.

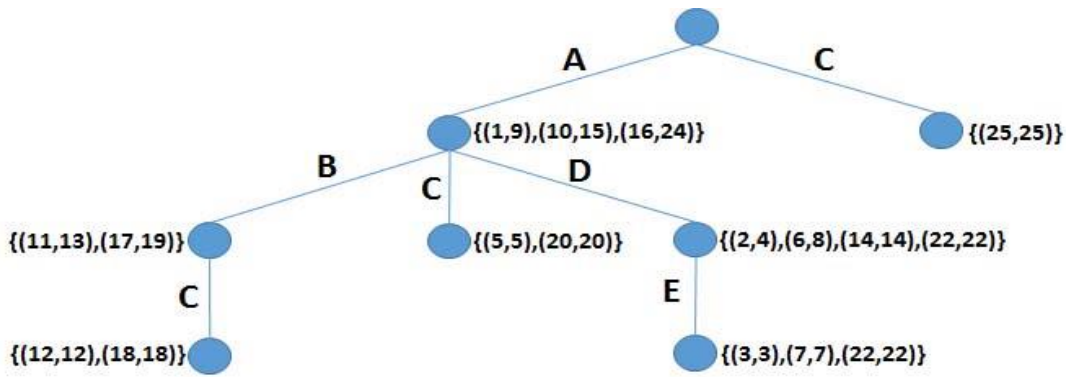

Figure 5. Annotated DataGuide (index) of the XML document of figure 4.c.

Following the presentation in section 3.2.1, the list associated with the query node *A/B!* is *TB* =[$\varepsilon_0$(1, 9), $\varepsilon_1$(10, 11), $b_1$, $\varepsilon_2$(13, 15), $\varepsilon_3$(15, 16), $\varepsilon_4$(16, 17), $b_2$] where, the $\varepsilon_i$, $i = 0,\ldots, 4$ are the *pseudo occurrences* of *B*. The lists associated with the respective nodes *A*, *C* and */A/D/E* are respectively: *TA* =[$a_1$, $a_2$, $a_3$], *TC* =[$c_1$, $c_2$, $c_3$, $c_4$, $c_5$], *TE* =[$e_1$, $e_2$, $e_3$].

Figure 6 shows the state of the various stacks and lists associated with the different nodes during matching. The table 1 presents the *preferenceTable* table obtained at the end of phase 2, from which the best answer $\{< a_3, b_2, c_3, d_3, e_3 >\}$ is selected as the only dominant tuple. In fact, it's the only tuple containing an occurrence of the preference node *D*.

Table 1. The preferenceTable.

| Candidates solutions | Preferences Nodes |
| --- | --- |
|  | B |
| $< a_1, c_1, d_1, e_1 >$ | 0 |
| $< a_1, c_1, d_2, e_2 >$ | 0 |
| *$< a_3, b_2, c_3, d_4, e_3 >$* | *1* |
| $< a_2, c_4, d_4, e_3 >$ | 0 |

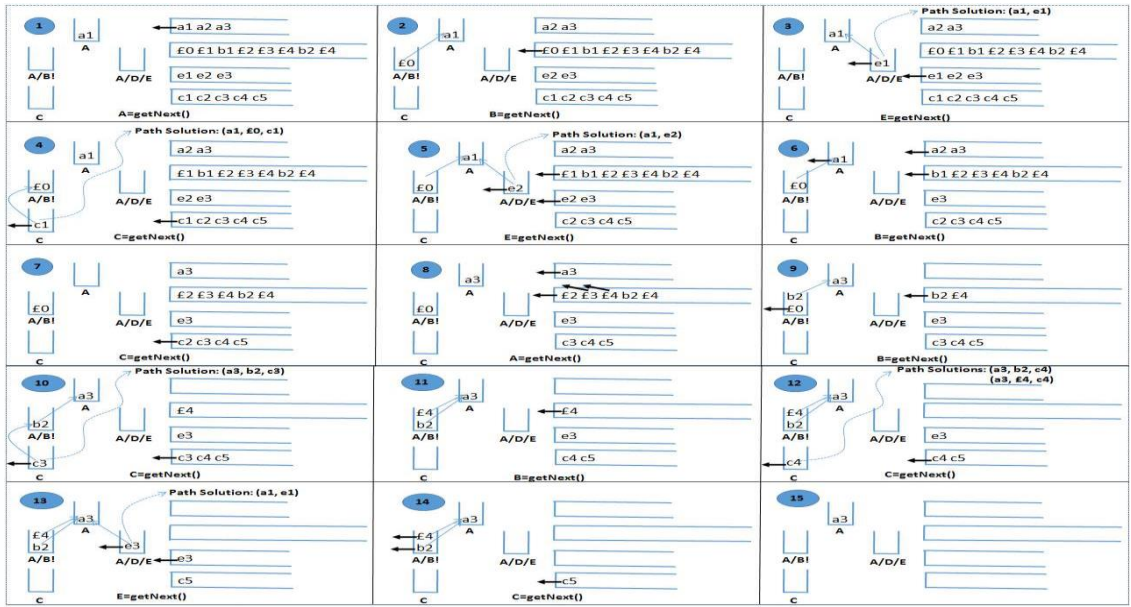

Figure 6. An illustration of the running of prefTwigStack: query of the figure 4.b is matching on the document of the figure 4.c.

## 5. CONCLUSIONS

We have presented in this paper a holistic evaluation approach of *XML* queries with structural preferences. This approach has been structured in three phases: rewriting-evaluation-merge. As in TwigX-Guide [4], a complex query is rewritten via a partitioning-transformation operation into a structured set of path queries that are holistically evaluated on an annotated DataGuide. The proposed approach reduces both the number of joins required for the holistic evaluation of a query, and the size of the lists to be joined.

The approach proposed in this manuscript has been carried out on many examples (among which the one presented in section 4) with very satisfactory results. With regard to the fact that the evaluation approach investigated in this paper is divided into three phases, the following question can be formulated: can deforestation techniques developed in functional programming be used to propose an evaluation algorithm that proceeds in a single phase, thereby avoiding the explicit construction of intermediate candidate responses? This seems to us to be the main object of an interesting study that can follow this work.